\documentclass[12pt,manuscript=article]{achemso}
\setkeys{acs}{maxauthors = 0}
\usepackage{graphics,epsfig,color,amsmath}
\usepackage{subcaption}
\usepackage{comment}
\title{Performance and Analysis of the Alchemical Transfer Method  for Binding Free Energy Predictions of Diverse Ligands}

\author{Lieyang Chen}
\affiliation{Roivant Sciences, 151 W 42nd Street, 15th Floor, New York, NY 10036, United States}
\author{Yujie Wu}
\affiliation{Roivant Sciences, 151 W 42nd Street, 15th Floor, New York, NY 10036, United States}
\alsoaffiliation{Atommap Corporation, New York, NY 10017}
\author{Chuanjie Wu}
\affiliation{Roivant Sciences, 151 W 42nd Street, 15th Floor, New York, NY 10036, United States}
\author{Ana Silveira}
\affiliation{Psivant Therapeutics, 451 D Street, Boston, MA 02210, United States}
\author{Woody Sherman}
\affiliation{Psivant Therapeutics, 451 D Street, Boston, MA 02210, United States}
\author{Huafeng Xu}
\affiliation{Roivant Sciences, 151 W 42nd Street, 15th Floor, New York, NY 10036, United States}
\alsoaffiliation{Atommap Corporation, New York, NY 10017}
\author{Emilio Gallicchio}
\email{egallicchio@brooklyn.cuny.edu}
\affiliation{Department of Chemistry, Brooklyn College of the City University of New York, New York, NY}
\alsoaffiliation{Ph.D. Program in Chemistry, The Graduate Center of the City University of New York, New York, NY}
\alsoaffiliation{Ph.D. Program in Biochemistry, The Graduate Center of the City University of New York, New York, NY}

\begin{document}

\maketitle

\begin{abstract}
The Alchemical Transfer Method (ATM) is herein validated against the relative binding free energies of a diverse set of protein-ligand complexes. We employed a streamlined setup workflow, a bespoke force field, and the AToM-OpenMM software to compute the relative binding free energies (RBFE) of the benchmark set prepared by Schindler and collaborators at Merck KGaA. This benchmark set includes examples of standard small R-group ligand modifications as well as more challenging scenarios, such as large R-group changes, scaffold hopping, formal charge changes, and charge-shifting transformations. The novel coordinate perturbation scheme and a dual-topology approach of ATM address some of the challenges of single-topology alchemical relative binding free energy methods. Specifically, ATM eliminates the need for splitting electrostatic and Lennard-Jones interactions, atom mapping, defining ligand regions, and post-corrections for charge-changing perturbations. Thus, ATM is simpler and more broadly applicable than conventional alchemical methods, especially for scaffold-hopping and charge-changing transformations. Here, we performed well over 500 relative binding free energy calculations for eight protein targets and found that ATM achieves accuracy comparable to existing state-of-the-art methods, albeit with larger statistical fluctuations. We discuss insights into specific strengths and weaknesses of the ATM method that will inform future deployments. This study confirms that ATM is applicable as a production tool for relative binding free energy (RBFE) predictions across a wide range of perturbation types within a unified, open-source framework.
\end{abstract}

\section{\label{sec:intro}Introduction}

Alchemical binding free energy prediction tools are emerging as a best-in-class standard for \textit{in silico} prediction of binding free energies (i.e., protein-ligand binding affinity) in structure-based drug design.\cite{abel2017advancing,armacost2020novel,zhang2021potent,Allen2022.05.23.493001,ganguly2022amber,xu2022slow} 
While many challenges remain,\cite{mobley2012let,lee2020alchemical} the increased utilization of RBFE calculations has been fueled in part by the promising results of large-scale validation campaigns against benchmark sets representative of actual drug discovery projects.\cite{GallicchioSAMPL4,wang2015accurate,zou2019blinded,schindler2020large,lee2020alchemical,kuhn2020assessment,gapsys2020large,bieniek2021ties,hahn2022bestpractices,gapsys2022pre,sabanes2023validation,cournia2017relative}

Despite decades of progress, reliable prediction of binding affinities of protein-ligand complexes remains a challenging problem with many unresolved issues, especially related to large chemical modifications, core transformations, and formal charge changes. In principle, the dissociation constant of a complex can be measured via brute force molecular dynamics (MD) simulations by sampling many binding/unbinding events.\cite{pan2017quantitative} However, this approach is generally not practical because the typical residence time of the ligand in the binding site (from milliseconds to hours) is too computationally expensive for practical applications in drug discovery. The alternative physical pathway methods\cite{Deng2009,gumbart2013efficient,velez2013overcoming,limongelli2013funnel,lapelosa2017free,deng2018comparing,cruz2020combining,azimi2022application} consist of measuring the reversible work of dragging the ligand from the solution to the binding site (or vice versa) along a chosen route. While physically appealing and computationally more efficient than brute force MD,\cite{comer2015adaptive,deng2018comparing,mahinthichaichan2021kinetics,tse2020exploring} physical pathway methods are rarely used in small molecule structure-based drug design because they require overcoming transition states and they do not readily apply to the common enclosed binding sites that lack a clear entryway.\cite{cruz2020combining} Furthermore, physical pathway methods do not apply to the direct estimation of relative binding free energies useful in drug discovery applications. 

In drug discovery applications, knowledge of the change in binding affinity resulting from modifying a ligand into another is usually more relevant than that of their absolute binding affinities due to the nature of the iterative design-make-test (DMT) cycle that is almost always employed to advance a hit to a development candidate.\cite{wang2015accurate,armacost2020novel,schindler2020large} Furthermore, while sampling of the full binding/unbinding pathway can yield information about binding kinetics, it is unnecessary for the computation of binding thermodynamics. Accordingly, Relative Binding Free Energy (RBFE) alchemical protocols can estimate the difference in the binding free energies of a pair of related ligands more directly than computing each of their absolute binding free energies.\cite{Jorgensen2004,cournia2017relative,Mey2020Best,azimi2022relative} That being said, absolute binding free energies still have great potential value in the context of virtual screening of diverse molecules.\cite{cournia2020rigorous} This work focuses on relative binding free energies in the context of optimizing screening hits to drugs. 

In most RBFE software implementations, the transformation of one molecule to another is accomplished by parameter interpolation approaches\cite{liu2013lead,Mey2020Best} that scale parameters of the potential energy function to convert one ligand into another through an alchemical transformation. However, parameter interpolation schemes require complex and often non-transferable customization of the energy subroutines of molecular dynamics engines and custom soft-core pair-potentials used to correct singularities of the alchemical potential energy function.\cite{Steinbrecher2011,lee2020improved} Moreover, parameter interpolation implementations typically do not directly connect the two ligands in their bound states. Rather, they rely on a thermodynamic cycle and alchemical calculations in solution and receptor environments separately, and often each step is further split into the decoupling of electrostatic and non-electrostatic interactions to avoid numerical instabilities.\cite{Mey2020Best,lee2020alchemical} Furthermore, transformations involving changes in the net ligand charge require the additional calculation of correction factors.\cite{dixit2001can,wallace2012charge,chen2013introducing,chen2018accurate,rocklin2013calculating}

Atom mapping procedures to find corresponding atom pairs for interpolation and the creation or annihilation of atoms to dummy types add complexities to current single-topology alchemical RBFE protocols.\cite{fleck2021dummy,zou2019blinded,jiang2019computing,Gallicchio2021binding}  Single-topology transformations often encode non-standard molecular topology formats that require custom system setup workflows that are incompatible with standard molecular visualization and trajectory analysis tools. In many implementations, RBFE alchemical schemes are limited to R-group transformations involving ligand pairs sharing a common scaffold.\cite{lee2020alchemical,zhang2021charmm} Except for a few commercial products,\cite{wang2017accurate,raman2020automated,zou2021scaffold} scaffold-hopping RBFE calculations involving cyclization, ring expansion, linking, or any other transformation that necessitate the breaking or the formation of chemical bonds,\cite{liu2015ring} are not generally supported.  

We have recently developed the Alchemical Transfer Method (ATM) that resolves many of the aforementioned complexities of RBFE calculations. The key innovations of the method are 1) the mapping of the potential energy functions of the unbound and bound states by a coordinate transformation rather than a variation of parameters and 2) expressing the alchemical potential energy function in a dual-topology formulation as the combination of the energy functions of the physical end states rather than the interpolation of their parameters.\cite{khuttan2021alchemical,wu2021alchemical,azimi2022application,azimi2022relative}

ATM is not as affected by the complexities of traditional alchemical methods. It supports absolute and relative binding free energy calculations in a unified way, computes free energies directly employing a single simulation box with standard chemical topologies, and natively supports diverse perturbations (standard R-groups, charge-changing, and scaffold-hopping transformations) without correction factors and ancillary calculations. Furthermore, since ATM employs a dual-topology formalism and does not use parameter interpolation or custom soft-core pair potentials, it is easier to implement and more straightforward to transfer across MD engines because it uses the unmodified energy routines of the underlying molecular dynamics engine. For the same reason, it applies to any molecular energy function, including the next generation of polarizable,\cite{harger2017tinker,panel2018accurate,Huang2018drude,das2022development} quantum-mechanical,\cite{beierlein2011simple,lodola2012increasing,hudson2019use,casalino2020catalytic} and machine-learning potentials\cite{smith2019approaching,rufa2020towards} that are starting to be employed in macromolecular simulations. The current open-source software release of ATM employs the OpenMM molecular dynamics engine and has been successfully tested on a series of drug discovery targets with the AMBER molecular mechanics force field in academic and industrial settings.\cite{azimi2022relative,sabanes2023validation}

The simplifications and greater range of applicability afforded by the alchemical transfer approach can be particularly useful in drug-discovery deployments to screen large and diverse ligand libraries in a more streamlined fashion. In this work, we validate ATM against the community benchmark prepared by Schindler et al.,\cite{schindler2020large} which contains examples of standard peripheral group transformations as well as more challenging scaffold-hopping and charge-changing transformations representative of real-world drug-discovery applications. 

Given the large number of calculations involved, aspects of the ATM workflow have been automated, which was facilitated by the nature of the ATM approach that avoids custom alchemical topologies and atom mapping typical of conventional RBFE workflows.\cite{lee2020alchemical,liu2013lead,Allen2022.05.23.493001,xu2019optimal,li2021precise}  Knowing that the quality of the potential energy functions can have a substantial effect on the free energy prediction accuracy,\cite{lu2021opls4,chung2023accurate} we applied a bespoke force field parameter generation protocol for each of the ligands. While the force field generation engine (FFEngine) is not publicly available, the parameters for each of the ligands in this work have been included in the Supporting Information. Thus, the work presented here is fully reproducible using the open-source version of ATM and the published force field parameters. 

The paper is organized as follows: We first introduce the theory of the Alchemical Transfer Method and present some of the key implementation details. We then describe the benchmarks sets, system setup, and alchemical simulation details. The results are presented and analyzed next. The paper concludes with a discussion of examples illustrating the strengths and weaknesses of the method and identifying areas of improvement for the application of ATM to drug discovery projects. 

\section{\label{sec:methods}Theory and Methods}

\subsection{The Alchemical Transfer Method (ATM)}

\begin{figure}
    \centering
    \includegraphics[scale=0.15]{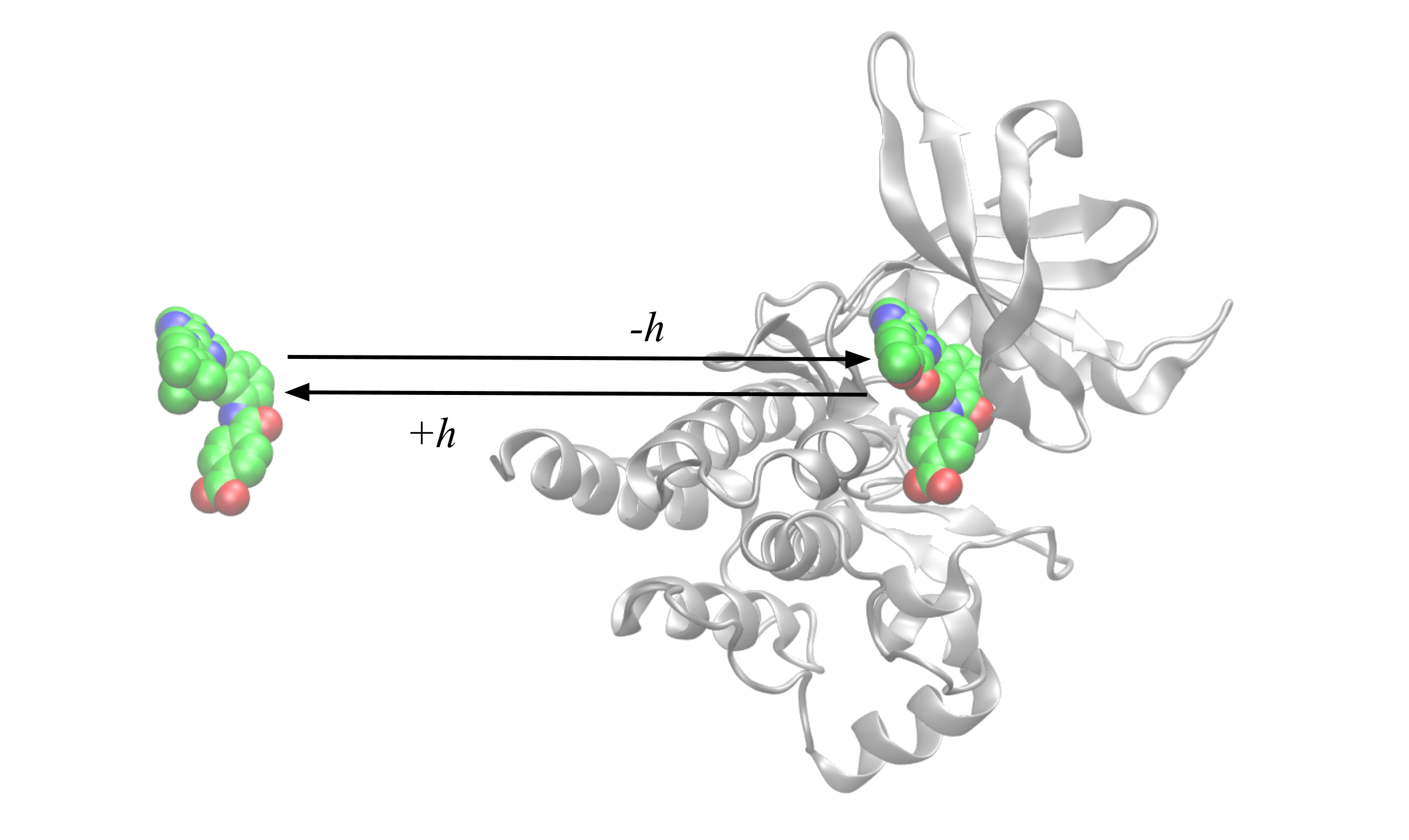}
    \caption{\label{fig:atm-cartoon} Illustration of the Relative Binding Free Energy calculation with the Alchemical Transfer Method. The first ligand is placed into the protein binding pocket, and the second in the solvent. During the alchemical transformation, the two ligands switch places. The resulting free energy corresponds to the difference in their binding free energies.}   
\end{figure}

The alchemical Transfer Method (ATM) models the free energy difference between two chemical states related by a coordinate transformation. One such example is the molecular binding processes investigated here, represented as the translation of the ligand from the solvent to the receptor binding site. The method details are fully described in previously published works.\cite{wu2021alchemical,azimi2022relative}, so an abridged account is provided here. Briefly, consider the potential energy, $U_0(x)$, of the unbound state ($R + A$) of the complex between a receptor R and a ligand A when the ligand is in solution and where $x = (x_R, x_A, x_S)$ represents the coordinates of the receptor, ligand, and solvent, respectively. ATM expresses the potential energy function, $U_1(x)$, that describes the state $RA$ when the ligand is bound to the receptor as
\begin{equation}
U_1(x) = U_0(x_R, x_A + h, x_S) \, ,
\label{eq:U1-atm-def}
\end{equation}
where $h$ is a fixed displacement vector that brings the ligand from its position in the solvent bulk to the receptor binding site (Figure \ref{fig:atm-cartoon}). The free energy difference between the bound and unbound states is then calculated by Free Energy Perturbation (FEP), similar to standard binding free energy methods. To this end, we define the perturbation energy function as
\begin{equation}
u(x) = U_1(x) - U_0(x) = U_0(x_R, x_A + h, x_S) - U_0(x_R, x_A, x_S)
\label{eq:upert-def}
\end{equation}
and introduce a $\lambda$-dependent alchemical potential energy function
\begin{equation}
U_\lambda(x) = U_0(x) + W_\lambda[u(x)] \, ,
\label{eq:Ulambda-def}
\end{equation}
where $W_\lambda(u)$ is an alchemical perturbation function with the properties $W_0(u) = 0$ and $W_1(u) = u$, ensuring that Eq.\ (\ref{eq:Ulambda-def}) interpolates from the initial unbound state $U_0(x)$ at $\lambda=0$ and the final bound state $U_1(x)$ at $\lambda=1$. The linear alchemical perturbation function $W_\lambda(u) = \lambda u$ is a common choice. In this work, we adopt a non-linear expression described in Computational Details that yields faster convergence than the default linear version.\cite{pal2019perturbation}

As elaborated elsewhere,\cite{wu2021alchemical,azimi2022relative} the alchemical path between the unbound and bound endpoints is divided into two legs: one starting from the unbound state at $\lambda = 0$ using the alchemical potential in Eq.\ (\ref{eq:Ulambda-def}), and a second leg starting from the bound state $U_1(x)$ morphing in the other direction towards the unbound state using the alchemical potential $U_\lambda(x) = U_1(x) + W_{1-\lambda}[-u(x)]$. Both legs terminate at $\lambda=1/2$ at the ATM alchemical intermediate with the potential energy function $U_{1/2}(x) = [U_0(x)+U_1(x)]/2$ that is an equally weighted average of the unbound and bound states. 

The ATM formulation above is for an Absolute Binding Free Energy (ABFE) calculation. Here, we are concerned with Relative Binding Free Energy (RBFE) prediction, in which the binding of a ligand $B$ occurs simultaneously as another ligand, $A$, leaves the receptor binding site. The free energy change of this process is the difference between the binding free energies of the two ligands. More specifically, an RBFE ATM calculation computes the free energy change from the state $RA + B$ with ligand $A$ bound to the receptor to the state $RB + A$ where ligand $B$ is bound to the receptor. In ATM, this process is described by a coordinate transformation that translates ligand $B$ from a position in the solvent to the binding site and simultaneously translates ligand $A$ from the receptor binding site to the solvent. Using the notation introduced above, the initial state $RA + B$ is described by the potential energy function $U_0(x)$, $x = (x_R, x_A, x_B, x_S)$, and the potential energy function $U_1(x)$ of the final state $RB + A$, where the positions of the two ligands are switched, is written as
\begin{equation}
U_1(x) = U_0(x_R, x_A - h, x_B + h, x_S) \, ,
\label{eq:U1-atm-rbfe-def}
\end{equation}
where the displacement vector $h$ translates ligand $B$ into the binding site while the opposite displacement $-h$ translates ligand $A$ in the reversed direction (Figure \ref{fig:atm-cartoon}). The nature of the coordinate transformation (a translation of one ligand vs.\ the translation of two ligands in opposite directions) is the only fundamental difference between ABFE and RBFE protocols in ATM. The definitions of the perturbation energy, the alchemical potential energy function [Eqs.\ (\ref{eq:upert-def})--(\ref{eq:Ulambda-def})] and the alchemical intermediate are the same. 

The main advantage of ATM for relative versus absolute binding free energy calculations is computational efficiency. For absolute binding free energy calculations it is necessary to converge the unbound state of the receptor, which can be a slow process given the possibility of high-barrier conformational rearrangements, differences in binding site solvent structure, changes to the protonation/tautomerization states of binding site residues, and other differences that may exist between the bound and unbound states. ATM RBFE calculations always have a ligand in the binding site, thus minimizing these effects. The actual free energy of the unbound receptor is inconsequential in the context of RBFE calculations because it is a constant for each ligand. 

The alchemical intermediates in RBFE calculations are unphysical states in which the ligands are present simultaneously in the receptor binding site and solution, each at a strength proportional to the coupling parameter $\lambda$. The more dissimilar the ligands, the more strained the conformations of the system required to accommodate the alchemical intermediate states. This characteristic is reflected in the high free energy barrier encountered at the alchemical midpoint for some ligand pairs. The height of the free energy barrier is used below as one of the proxies to judge the quality of RBFE calculations and the confidence level of their estimates. The specific settings of the RBFE protocol used in this work are described in Computational Details.

\subsubsection{FFEngine Ligand Force Field Assignment}

FFEgine is a force field toolkit built with ParmEd\cite{davidcase2017parmed}, RDKit\cite{rdkit}, GAFF2\cite{wangj2020gaff2}, GFN2-xTB\cite{grimme2019xtb}, GPU-powered QM package Terachem\cite{martinez2009, martinez2020} that provides high-quality parameters for drug-like molecules based on a bespoke workflow where quantum mechanical calculations are performed on each molecule to obtain the potential surface for force field fitting. The atom types in FFEngine are defined following the hierarchical structure described by Jin Z.\ et al.\cite{huaisun2016} In total, FFEngine utilizes approximately 200 atom types, with the objective of covering the chemical space of drug-like molecules. 

FFEngine generates ligand parameters for the AMBER force field functional form\cite{wang2006automatic}, for use in AMBER, GROMACS, OpenMM, or other software packages. The charge assignment uses GFN2-xTB/BCC, which is similar to the AM1/BCC model from AMBER. Atomic partial charges are assigned based on atom types and the bond charge correction (BCC) parameters fitted to the HF/6-31G* electrostatic potential (ESP) from 50,000 drug-like compounds and their conformations. In GFN2-xTB/BCC, the pre-charges are assigned with the semiempirical method GFN2-XTB. The molecular structures are relaxed with the machine learning force field GFN-FF to prepare the structure before partial charges assignment.\cite{grimmer2020gfn-ff} The vdW, bond, angle, and torsion parameters are assigned based on a more refined set of atom types. GAFF2 was used as the fallback parameters for bond, angle, and torsional degrees of freedom.

\subsubsection{DiffNet Analysis}

ATM yields estimates of the differences between the binding free energies of pairs of ligands. We employed DiffNet\cite{xu2019optimal,DiffnetRepository} to estimate the binding free energies of the ligands in the set. Diffnet finds the set of binding free energies most consistent with the network of their differences (referred to as {\em edges}) and their uncertainties based on the maximum likelihood principle. The DiffNet solution is known up to an arbitrary constant that we set to match the average of the experimental binding free energies.

The uncertainties associated with the edges of the network of free energy transformations are an important element of the DiffNet protocol. Edges with small uncertainties weigh on the final solution more than edges with larger uncertainties. To define confidence levels in the ATM RBFE predictions, we used two measures of the quality of the alchemical calculations: 1) gaps in the perturbation energy distributions and 2) height of the free energy barrier along the alchemical path.

To assess the presence of gaps in the perturbation energy distributions, the binding energy samples of each leg were histogrammed with bins of size $\Delta u = k_B T/\Delta \lambda$, where $\Delta \lambda = 1/N_{\rm states}$ is the size of the subdivision of the alchemical path among $N_{\rm states}$ alchemical states (approximately $13$ kcal/mol in this application). A case with a sequence of two or more bins with zero counts was flagged as unlikely to be converged, and its uncertainty was increased by $\Delta u$. A case with a sequence of two or more bins with less than 10\% of the expected number of samples based on a uniform distribution was flagged as possibly unconverged, and its uncertainty was increased by a tenth of $\Delta u$. Similarly, to capture the lower confidence of predictions of large transformations with high free energy intermediates, we increased the uncertainty of predictions where the free energy of the alchemical intermediate, $\Delta \Delta G(1/2)$, exceeds $30$ kcal/mol relative to either end state. Specifically, in these cases, we increased the uncertainty linearly by $ a (\Delta \Delta G(1/2) - g)$ where $a = 0.1$ kcal/mol and $g = 30$ kcal/mol. While this assessment successfully flagged possibly problematic calculations, the specific parameters used here to assign confidence levels were set empirically and would benefit from validation in future studies.

\subsubsection{Simulation Setup Workflow}

To conduct this work, we wrote a Python-based ATM RBFE setup and analysis workflow. The workflow performs  water placement,  force field generation, reference atom selection, displacement vector searching, AToM-OpenMM simulation, relative binding free energy calculation, and DiffNet analysis. As input, the workflow requires fully prepared proteins and docked ligand poses. 

 In order to get better solvation structures for the receptors, 3D-RISM\cite{stumpe2011calculation} from AmberTools\cite{AmberTools} was first applied to estimate the implicit water distribution on the receptor-primary ligand complex surface, followed by Placevent\cite{Placevent} to place the explicit water molecules on the complex surface. Any water molecules that clashed with other ligands in the set were removed. The workflow then assigned ligand force field parameters using FFEngine as described above.  

An ATM RBFE calculation requires the choice of three corresponding reference atoms for the alignment of the ligand pair.\cite{azimi2022relative} The alignment reference atoms cannot be colinear and are best chosen among rigid core atoms of the two ligands. This task involves manual selection the reference atoms of only one representative ligand placed into the receptor binding site. All other ligands are then automatically aligned to the first by a minimum distance search based on their initial poses.

In the simulation box for an ATM calculation, one ligand is placed in the binding pocket, and the other one is translated into the solvent phase by a displacement vector $h$. While the choice of the displacement vector could be arbitrary, a reasonable choice should ensure that any atom of the second ligand is at least 10 \AA\ away from any of the receptor atoms. On the other hand, a large displacement would unnecessarily increase the box size, affecting performance. We used TLeap\cite{AmberTools} to generate a preliminary rectangular solvation box for the receptor with a 10 \AA\ water buffer to select a good displacement vector automatically. The ATM displacement vector $h$ was then obtained by displacing the ligand from the binding pocket towards the center of the box face with the smallest surface area until all atoms of the ligand were found outside the box.

The Amber input chemical topologies with the assigned force field parameters generated by our automated workflow are listed in the simulation input files available on the GitHub repository at {\tt https://github.com/EricChen521/ATM\_MerckSet}.

\subsection{Benchmark Systems}

The benchmark sets prepared by Schindler et al.\cite{schindler2020large} include eight receptor targets with 24 to 44 ligands for each target (Table \ref{tab:stats}). The benchmark also specifies the ligand pairs forming the edges of the graph of RBFE calculations.\cite{SchindlerRepository} In total, the benchmark set includes 264 protein-ligand complexes and 550 RBFE edges. Schindler et al.\ reported free energy estimates for 525 of the 550 transformations using Schr\"{o}dinger's FEP+ package.\cite{wang2015accurate} In this work, we considered all of the protein-ligand complexes and the corresponding edges except the two involving compound 28 of the CDK8 set, which we suspect to have been misidentified. This compound lacks a measured binding free energy and is classified as a non-binder, even though close analogs are strong binders.\cite{schiemann2016discovery}

The benchmark sets provided by Schindler et al.\cite{schindler2020large} are considered challenging because, in addition to standard small R-group modifications, they include a significant number of more challenging transformations. The set includes 43 large R-group transformations (more than 10 added/removed atoms), 66 charge transformations (either changing the formal charge or moving the location of the formal charge), and 60 scaffold-hopping transformations. Large R-group transformations are considered challenging because they often induce changes in the conformation and hydration pattern of the complex to accommodate the new groups of atoms. Charge-changing RBFEs, which involve ligands with different net charges, have traditionally required specialized strategies or correction terms\cite{pan2017quantitative,ohlknecht2020correcting,rocklin2013calculating} unnecessary in our alchemical transfer approach.\cite{azimi2022relative} Nevertheless, charge-changing transformations and the related charge-shifting transformations remain challenging because of the receptor and solvent reorganization that they often induce. Finally, scaffold-hopping transformations that include cyclization, ring-breaking, and ring expansion/reduction transformations that involve the breaking or forming of chemical bonds, which traditionally require specialized strategies,\cite{wang2017accurate,zou2021scaffold} are generally straightforward and are treated here with ATM as any other transformation.\cite{azimi2022relative}

\subsection{Computational Details}

\subsection{Molecular Systems Setup}

We employed the structures of the proteins and bound ligands as provided by Schindler et al.\cite{schindler2020large} The simulation systems were prepared for the RBFE calculations using the automated workflow described above. The AMBER FF14SB force field\cite{maier2015ff14sb} was used for the protein and the TIP3P model\cite{jorgensen1983comparison} for water. K$^+$ and Cl$^-$ ions were added to the system if needed for neutralization. The force constants of the ligand alignment restraints were set to $k_r = 2.5$ kcal/(mol \AA$^2$) for the position restraint, and $k_\theta = k_\psi = 50$ kcal/mol for the orientational restraints.\cite{azimi2022relative} The C$\alpha$ atoms' positions of the receptors were restrained to their initial values using flat-bottom harmonic restraints with a tolerance of  $1.5$ \AA\ and a force constant of $25$ kcal/(mol \AA$^2$).

The solvated systems were energy minimized, thermalized, and relaxed for 400 ps at 298 K and 1 bar constant pressure and annealed to the $\lambda = 1/2$ alchemical intermediate in 500 ps while restraining the receptor and the ligands' atoms. The system was then equilibrated in the NVT ensemble at 298 K for 300 ps at the alchemical intermediate after releasing the restraints on the ligands and the receptor (except for the C$\alpha$ atoms positional restraints described above). The resulting structures were used as starting configurations for the alchemical replica exchange simulations described next. 

\subsection{Alchemical Transfer Relative Binding Free Energy Protocol}

ATM RBFE calculations were conducted using the AToM-OpenMM package version 3.2.3,\cite{AToM-OpenMM} the ATM MetaForce OpenMM plugin version 0.3.1,\cite{ATMMetaForce-OpenMM-plugin} and the OpenMM MD engine version 7.7.\cite{eastman2017openmm}. The ATM alchemical schedule is comprised of two legs.\cite{wu2021alchemical,azimi2022relative} Using the notation introduced in Theory and Methods, the first leg corresponds to the transformation from the $RA + B$ state described by the potential energy function $U_0(x)$ to the alchemical intermediate state $R(AB)_{1/2} + (AB)_{1/2}$ described by the potential energy function $U_0(x)/2 + U_1(x)/2$. The second leg reaches the same alchemical intermediate state but starts from the final $RB + A$ state. The difference between the free energy changes of the first leg and second legs yields the relative binding free energy $\Delta\Delta G_b^\circ(B,A)$ of ligand B with respect to ligand A. The alchemical potential energy function for the first leg is given in Eq.\ (\ref{eq:Ulambda-def}) while that for the second leg is
\begin{equation}
U_{1-\lambda}(x) = U_1(x) + W_\lambda[-u(x)] \, ,
\label{eq:Ulambda-def-back}
\end{equation}
where $0 \le \lambda \le 1/2$, the perturbation energy is given in Eq.\ (\ref{eq:upert-def}), and, for both legs, the alchemical perturbation function is the softplus function\cite{pal2019perturbation}
\begin{equation}
  W_{\lambda}(u)=\frac{\lambda_{2}-\lambda_{1}}{\alpha}\ln\left[1+e^{-\alpha(u_{\rm sc}(u)-u_{0})}\right]+\lambda_{2}u_{\rm sc}(u) .
  \label{eq:ilog-function}
\end{equation}
The parameters $\lambda_{2}$, $\lambda_{1}$, $\alpha$, and $u_{0}$ are functions of $\lambda$ (see below).\cite{khuttan2021alchemical}. The function
\begin{equation}
  u_{\rm sc}(u)=
\begin{cases}
u & u \le u_c \\
(u_{\rm max} - u_c ) f_{\rm sc}\left[\frac{u-u_c}{u_{\rm max}-u_c}\right] + u_c & u > u_c
\end{cases}
\label{eq:soft-core-general}
\end{equation}
with
\begin{equation}
f_\text{sc}(y) = \frac{z(y)^{a}-1}{z(y)^{a}+1} \label{eq:rat-sc} \, ,
\end{equation}
and
\begin{equation}
    z(y)=1+2 y/a + 2 (y/a)^2
\end{equation}
with, in this work, $u_{\rm max} = 200$ kcal/mol, $u_c = 100$ kcal/mol, and $a = 1/16$, is the soft-core perturbation energy function designed to avoid singularities near the initial state of the alchemical transformation.\cite{khuttan2021alchemical}

A schedule of 11 equispaced $\lambda$-states from $\lambda=0$ to $\lambda=1/2$ was employed for each of the two legs. The corresponding schedules of softplus alchemical parameters in Eq.\ (\ref{eq:ilog-function}) were: $\lambda_1 = 0, 0, 0, 0, 0, 0, 0.1, 0.2, 0.3, 0.4, 0.5$, $\lambda_2 = 0, 0.1, 0.2, 0.3, 0.4, 0.5, 0.5, 0.5, 0.5, 0.5, 0.5$, and $\alpha = 0.1$ (kcal/mol)$^{-1}$ and $u_0 = 110$ kcal/mol for all $\lambda$-states. We employed this alchemical schedule for all the ligand pairs of the benchmark sets in this work without further optimization.

For each run, asynchronous Hamiltonian replica exchange\cite{gallicchio2015asynchronous} molecular dynamics conformational sampling  was performed with a 2 fs timestep and a replica running time of 40 ps on 4 GPUs using the AToM-OpenMM software.\cite{AToM-OpenMM} Exchanges between the two equivalent alchemical intermediate states at $\lambda=1/2$ allow replicas to transition from one alchemical leg to the other. Perturbation energy samples were collected every 40 ps. Relative binding free energies were computed from replica trajectories at least 5 ns long (approximately 110 ns of MD in aggregate per ligand pair), discarding the first third of the samples for equilibration. UWHAM multi-state analysis\cite{Tan2012} was used for free energy and statistical error estimation. 

The molecular system files, AToM-OpenMM input files, and UWHAM analysis code are available on GitHub at {\tt https://github.com/EricChen521/ATM\_MerckSet}.

\section{Results}

\begin{figure}
    \centering
    \includegraphics[scale = 0.65]{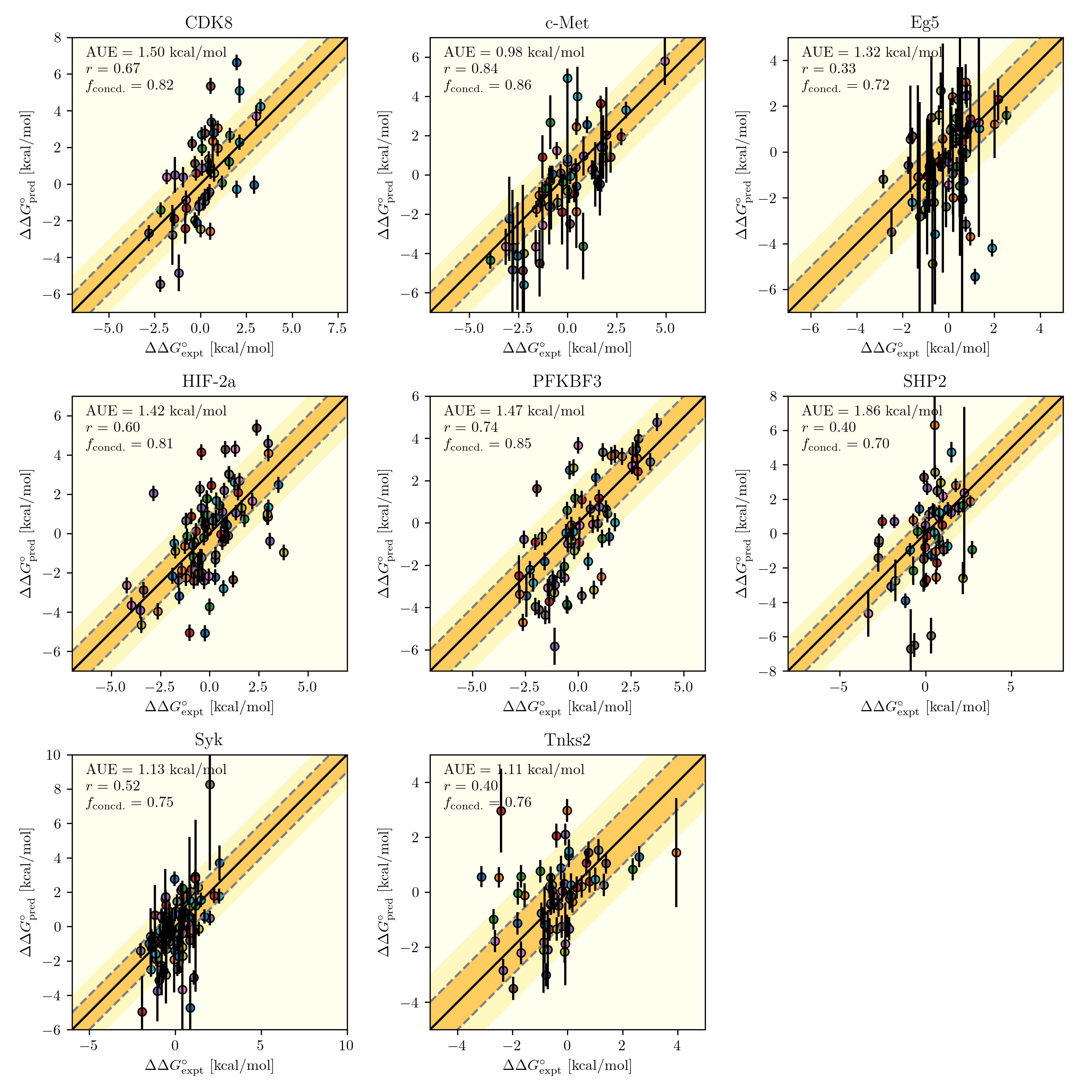}
    \caption{Scatter plots of the calculated RBFEs vs.\ the corresponding differences of experimental binding free energies. The 1:1 line represents perfect agreement. The dark and light-shaded areas correspond to predictions within 1 kcal/mol and 2 kcal/mol of the experiments, respectively. For each set the Average Unsigned Error (AUE), Pearson's correlation coefficient ($r$), and the fraction of concordant predictions ($f_{\rm concd.}$) are reported. }
    \label{fig:edges-corrplots}
\end{figure}

The relative binding free energy performance of the automated ATM protocol compared with the experimental measurements of the benchmark sets is summarized in Figure \ref{fig:edges-corrplots}, where each point represents a ligand pair. For each target set, we report the Average Unsigned Error (AUE), the Pearson's correlation coefficient $r$, and the fraction of concordant predictions $f_{\rm concd.}$ relative to the experiments. A concordant prediction is a case in which the direction of the change in binding affinity of more than $0.5$ kcal/mol is correctly predicted (see Methods). 

According to these statistics, the c-Met, PFKFB3, CDK8, and Hif-2$\alpha$ predictions are better than the other targets, with correlation coefficients of 0.6 or higher with over 80\% of concordant pairs. The performance for the c-Met set with $r=0.84$ and $f_{\rm concd.} = 86$\%  is particularly encouraging. However, due to a relatively small number of outliers, the AUEs for these sets are quite high (from $0.98$ to $1.5$ kcal/mol). The RBFE prediction performance for the remaining four sets (Eg5, SHP2, Syk, and TNKS2) is not as good. The Syk and TNKS2 set have fewer major outliers and display relatively small AUEs. However, the calculated RBFEs are poorly correlated with the experiments with correlation coefficients of 0.52 and 0.40, respectively. The SHP2 and, particularly, the Eg5 sets have both a significant number of outliers and poor correlation with the experiments.

The confidence levels of the RBFE predictions represented by the error bars in Figure \ref{fig:edges-corrplots} vary significantly from set to set. As described in the Methods, confidence levels reflect the difficulty of the alchemical calculation. They are assigned based on statistical fluctuations, connectedness, and the height of the free energy barrier of the alchemical pathway. Surprisingly, the calculations for c-Met, which are the closest to the experiments, generally have higher uncertainty than the other sets. The RBFE calculations for the Eg5 and Syk sets are also considered challenging. With a few exceptions, the RBFEs for the other sets are assigned with high confidence. The types of RBFE transformations and their challenges are further explored in the Discussion.

\begin{figure}
    \centering
    \includegraphics[scale = 0.65]{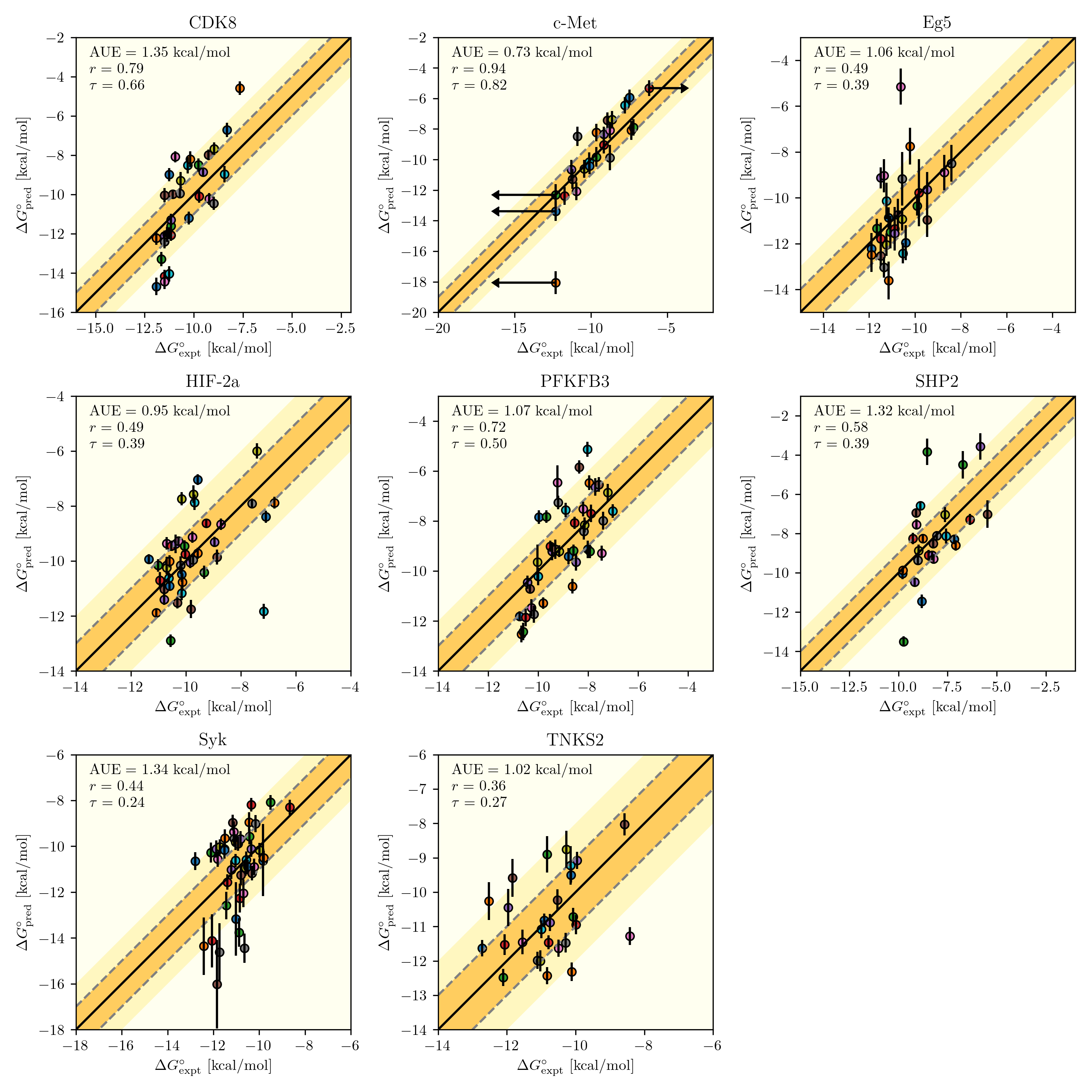}
    \caption{Scatter plots of the Diffnet ABFE estimates from the ATM RBFE values vs.\ the corresponding experimental binding free energies. The 1:1 line represents perfect agreement. The horizontal directed lines denote ligands for which the experimental affinity is known only within a range. The dark and light-shaded areas correspond to predictions within 1 kcal/mol and 2 kcal/mol of the experiments, respectively. For each set, the Average Unsigned Error (AUE), Pearson's correlation coefficient ($r$), and the Kendall's $\tau$ rank order correlation coefficient are reported.}
    \label{fig:corrplots}
\end{figure}

The accuracy of the absolute binding free energies (ABFE) predictions calculated from the RBFEs using the Diffnet protocol (see Methods) is summarized in Table \ref{tab:stats} and illustrated in Figure \ref{fig:corrplots}. The Pearson's correlation coefficients, Kendall's rank order correlation coefficients, and average unsigned errors relative to the experimental binding affinities averaged over the eight systems are $r=0.60$, $\tau=0.46$, and AUE $=1.11$ kcal/mol, respectively. These overall statistical measures are encouraging, given the unsupervised nature of the ATM calculation workflow and the challenges of the benchmarks. Results are slightly inferior to those obtained for the same benchmark set with the FEP+ software package ($r=0.62$, $\tau=0.50$, and AUE $= 0.98$ kcal/mol).\cite{schindler2020large}

Interestingly, the triangulation of the data afforded by DiffNet results in ABFEs significantly tighter agreement with the experiments than the RBFEs estimates. This is particularly noticeable for the c-Met set, which, despite the major outliers and large uncertainties of the RBFEs (Figure \ref{fig:edges-corrplots}),  yields high-confidence predictions with low AUEs (Figure \ref{fig:corrplots}).  

The ATM ABFE predictions for three of the eight systems (c-Met, PFKBF3, and CDK8) have high correlation coefficients and generally low AUEs relative to the experiments. FEP+, based on a different alchemical approach, performed similarly well for these systems, suggesting that they are tractable for diverse free energy methods.  Although affected by a few major outliers, the performance on the Hif-2$\alpha$ set is also generally good. Conversely, ATM performed noticeably worse for the SHP2 and Syk sets, where it resulted in an unusually poor AUE relative to the experiments. ATM and FEP+ performed equally poorly for the Syk and Eg5 systems indicating a common set of challenges. The TNKS2 set is less informative due to the small range of experimental affinities (Figure \ref{fig:corrplots}).

\begin{table*}
\begin{center}
\caption{\label{tab:stats} Statistical measures of the agreement between experimental and computed binding free energy values obtained with the ATM method compared with literature results using FEP+.\cite{schindler2020large}}
\begin{tabular}{lccccccccc}
Target   &     &&  \multicolumn{3}{c}{ATM}  && \multicolumn{3}{c}{FEP+\cite{schindler2020large}} \\
         &   $N$$^a$ &&    $r$$^b$ & $\tau$$^c$ & AUE$^d$      &&    $r$$^b$ & $\tau$$^c$ & AUE$^d$  \\
\hline
CDK8           & 32 &&   0.79 & 0.66 & 1.35   &&    0.62 & 0.57 & 1.20 \\ 
c-Met          & 24 &&   0.94 & 0.82 & 0.73   &&    0.90 & 0.73 & 0.82 \\
Eg5            & 28 &&   0.49 & 0.39 & 1.06   &&    0.47 & 0.54 & 1.09 \\
HIF-2$\alpha$  & 42 &&   0.49 & 0.39 & 0.95   &&    0.61 & 0.45 & 0.84 \\
PFKFB3         & 40 &&   0.72 & 0.50 & 1.07   &&    0.79 & 0.60 & 1.09 \\
SHP-2          & 26 &&   0.58 & 0.39 & 1.32   &&    0.71 & 0.61 & 0.74 \\
Syk            & 44 &&   0.44 & 0.24 & 1.34   &&    0.50 & 0.29 & 0.85 \\
TNKS2          & 27 &&   0.36 & 0.27 & 1.02   &&    0.40 & 0.29 & 1.23 \\
\hline
Cumulative & 263 &&   0.60 & 0.46 & 1.11   &&    0.62 & 0.50 & 0.98 \\
\hline
 \end{tabular}
 \end{center}
 $^a$ Number of complexes. $^b$ Pearson's correlation coefficient. $^c$ Kendall's rank order correlation coefficient. $^d$ Averaged unsigned error in kcal/mol.
\end{table*}

\section{Discussion}

The results obtained in this validation study confirm the applicability of the ATM approach to large-scale binding free energy estimation campaigns for challenging and diverse ligand libraries. In this section, we present some examples illustrating the strengths and weaknesses of the method that we observed during this work. We plan to incorporate these insights in deploying this approach in future work. 

\subsection{Small Charge-Preserving R-group RBFE Estimates are Generally Well-Converged}

\begin{figure}
    \centering
    \includegraphics[scale = 0.25]{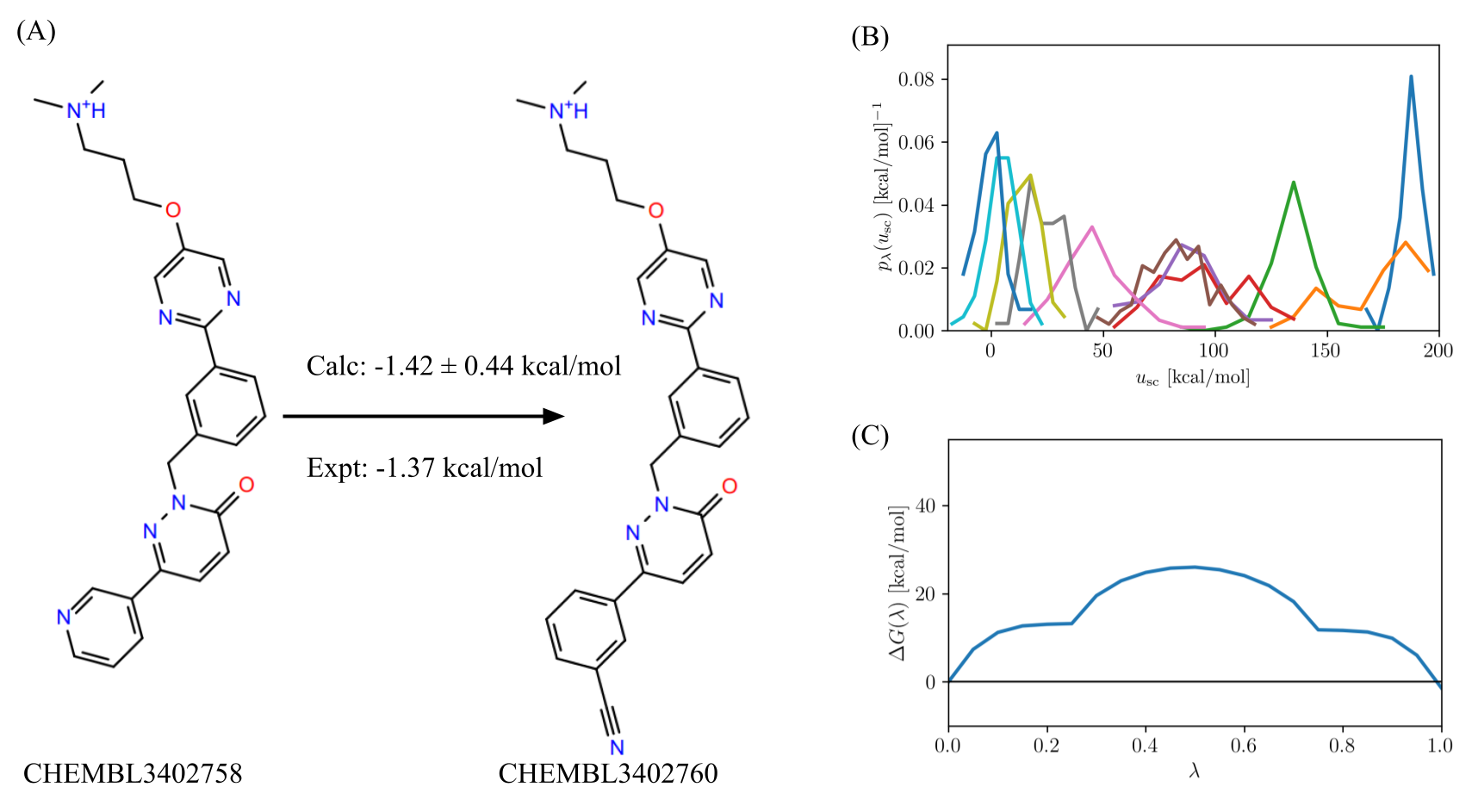}
    \caption{The results of the RBFE calculation between the CHEMBL3402758 and CHEMBL3402760 ligands of the c-Met receptor illustrating a reliable small R-group transformation. (A) Chemical structures of the ligands and calculated and experimental RBFE. (B) Perturbation energy distributions for the alchemical states from $\lambda=0$ to $\lambda = 1/2$ (from right to left in color sequence) showing good overlaps between nearby distributions. (C) Free energy profile along the alchemical transformation displaying a moderate free energy barrier.}
    \label{fig:cmet-CHEMBL3402758-10_CHEMBL3402760-1}
\end{figure}

Over 70\% of the ligand pairs of the benchmark (384 out of 548) are classified as conventional R-group alchemical transformations involving one or more small peripheral substitutes of the same net charge with 10 or fewer heavy atoms. The calculations of these pairs are expected to converge rapidly, and the corresponding relative binding free energy estimates to be more reliable than the other more challenging cases. As illustrated by the example in Figure \ref{fig:cmet-CHEMBL3402758-10_CHEMBL3402760-1}, this expectation is largely confirmed by the consistently good quality measures in terms of the overlaps between perturbation energy distributions and the moderate height of the free energy profile of R-group transformations. The RBFE estimates of R-group transformations are also generally in closer agreement and have fewer outliers relative to the experiments than other transformation types. For example, the RMSE of the R-group transformations for the CDK8, c-Met, and HIF-2$\alpha$ sets are $1.68$, $1.21$, and $1.53$ kcal/mol, respectively, compared to $1.87$, $1.38$, and $1.81$ kcal/mol for the entire sets. 

\begin{figure}
    \centering
    \includegraphics[scale = 0.25]{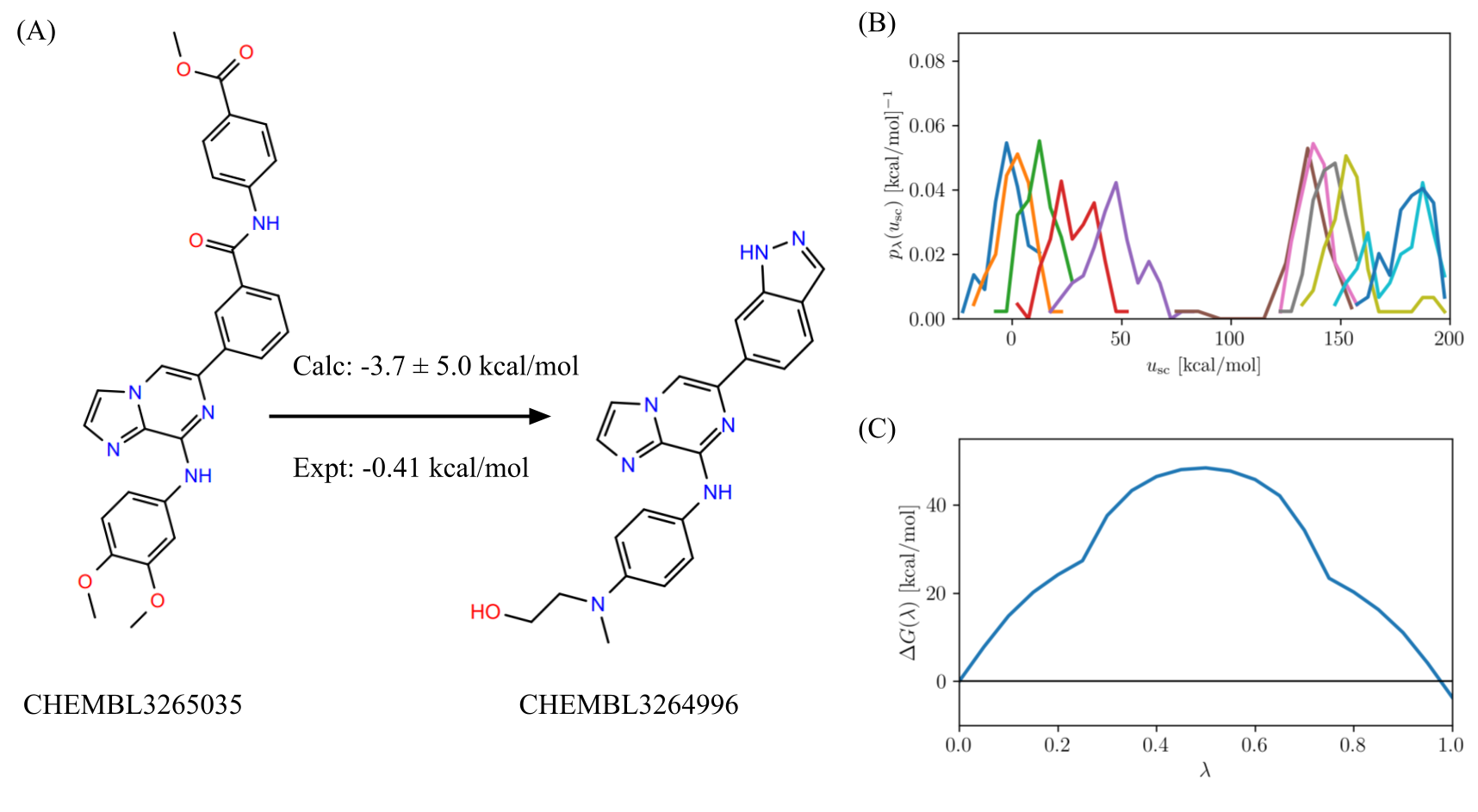}
    \caption{The results of the RBFE calculation between the CHEMBL3265035 and CHEMBL3264996 ligands of the Syk receptor illustrating an unreliable large R-group transformation. (A) Chemical structures of the ligands and calculated and experimental RBFE. (B) Perturbation energy distributions for the alchemical states from $\lambda=0$ to $\lambda = 1/2$ (from right to left in color sequence) showing a large gap between high-energy and low-energy distributions. (C) Free energy profile along the alchemical transformation displaying a large free energy barrier.}
    \label{fig:syk-CHEMBL3265035_CHEMBL3264996}
\end{figure}

Conversely, as illustrated by the Syk pair in Figure \ref{fig:syk-CHEMBL3265035_CHEMBL3264996}, RBFE estimates between very dissimilar ligands, even if involving only peripheral groups without charge variations, are often deemed unconverged and unlikely to be predictive of the actual difference in binding affinity. We generally observe this behavior when R-group modifications involve more than 10 heavy atoms.  These cases (43 out of 548) often present gaps in the sequence of perturbation energy distributions along the alchemical path and display a free energy barrier of the free energy profile of 40 kcal/mol or more (Figure \ref{fig:syk-CHEMBL3265035_CHEMBL3264996}). Many of the transformations in this category are outliers compared to the experiments. For example, nearly half of the pairs of the Syk set in this category deviate from the experiments by more than 2 kcal/mol. Transformations between very dissimilar ligands such as this are probably best implemented as sequences of smaller and more manageable alchemical steps.  

\subsection{Charge-changing and charge-shifting Transformations are Challenging}

\begin{figure}
    \centering
    \includegraphics[scale = 0.25]{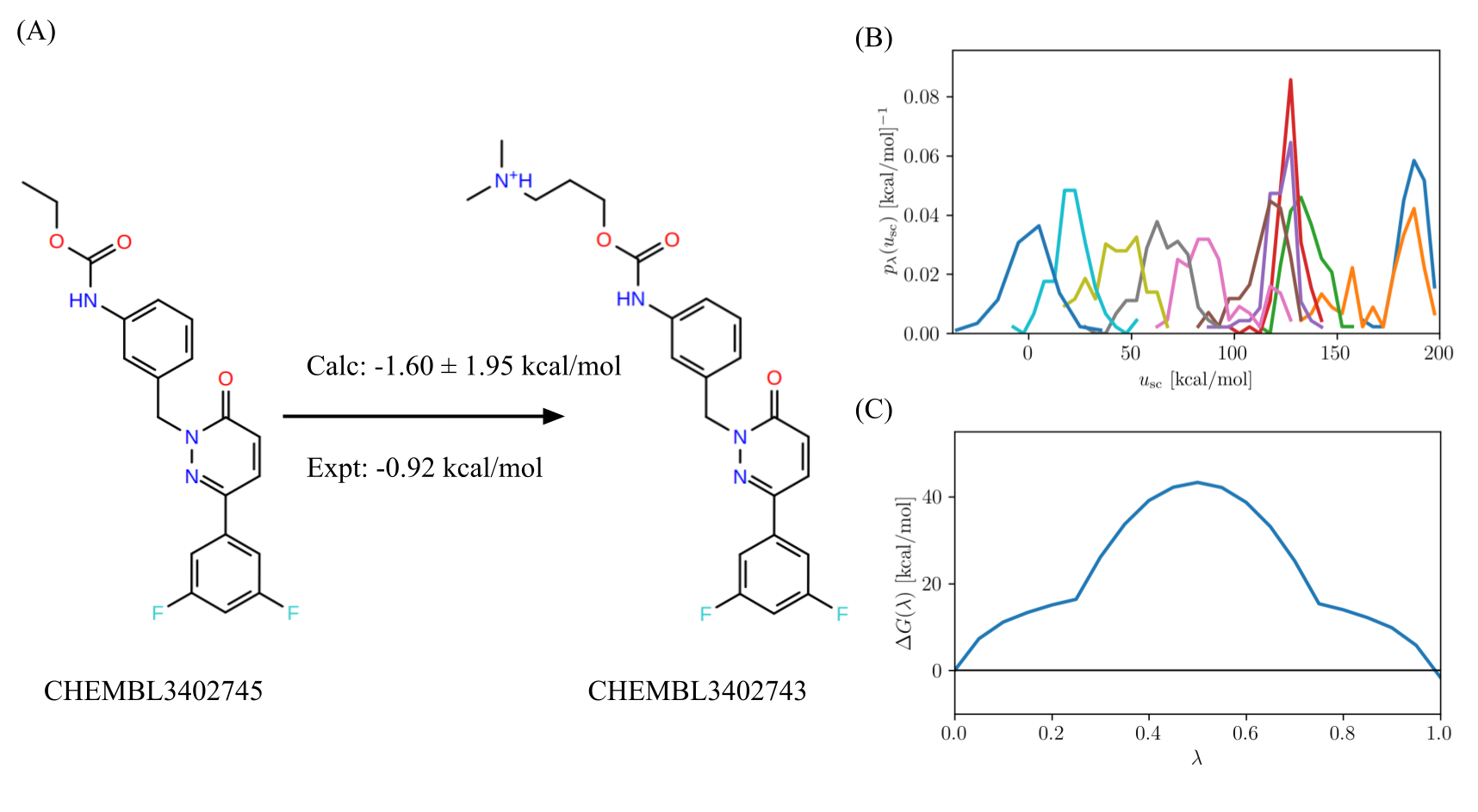}
    \caption{The results of the RBFE calculation between the CHEMBL3402745 and CHEMBL3402743 ligands of the c-Met receptor illustrating a possibly reliable charge-changing transformation. (A) Chemical structures of the ligands and calculated and experimental RBFE. (B) Perturbation energy distributions for the alchemical states from $\lambda=0$ to $\lambda = 1/2$ (from right to left in color sequence) showing mostly good overlaps but an undersampled region near $u_{\rm sc} \simeq 160$ kcal/mol. (C) Free energy profile along the alchemical transformation displaying a large free energy barrier.}
    \label{fig:cmet-CHEMBL3402745-200_CHEMBL3402743-42}
\end{figure}

\begin{figure}
    \centering
    \includegraphics[scale = 0.25]{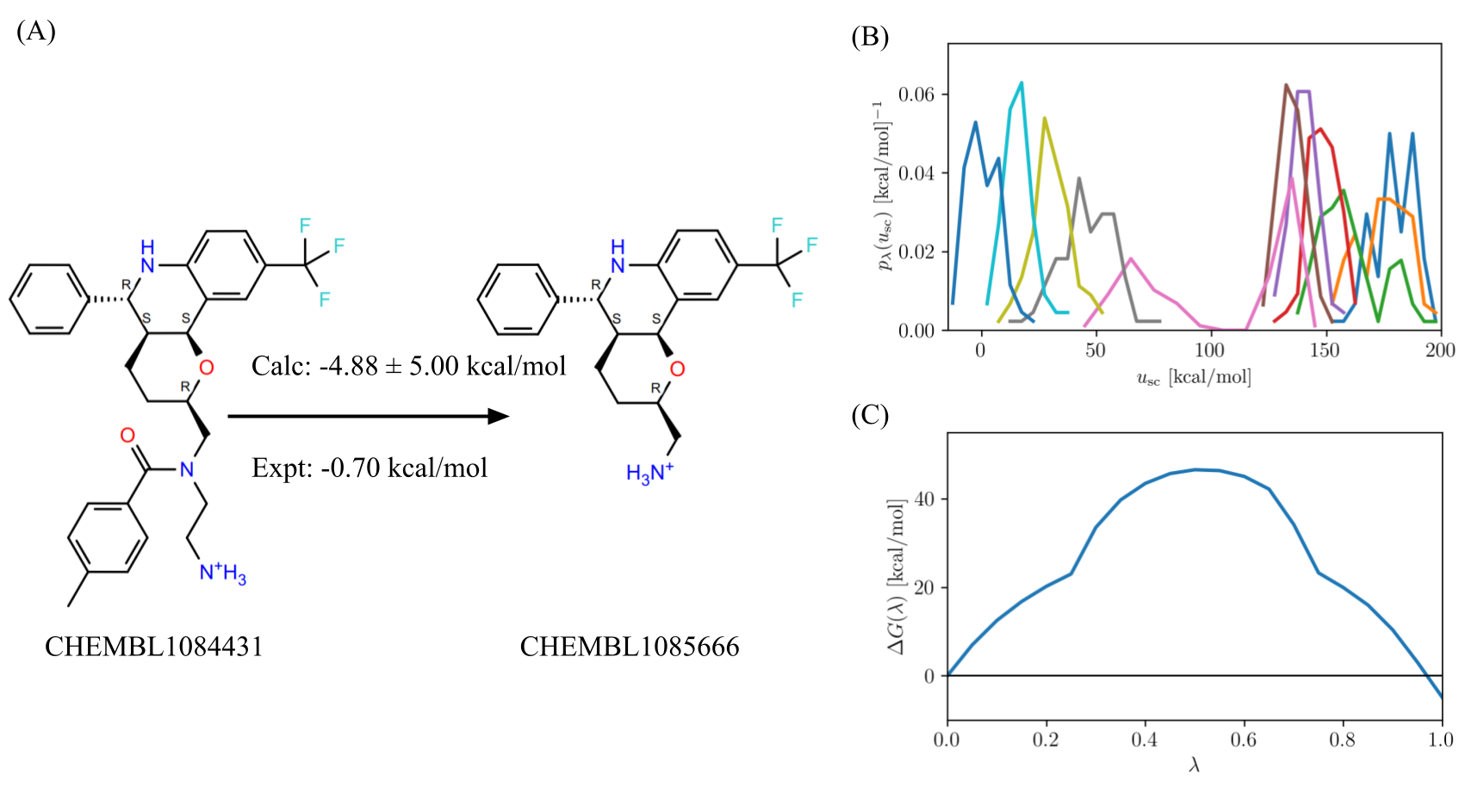}
    \caption{The results of the RBFE calculation between the CHEMBL1084431 and CHEMBL1085666 ligands of the Eg5 receptor illustrating an unreliable charge-shifting transformation. (A) Chemical structures of the ligands and calculated and experimental RBFE. (B) Perturbation energy distributions for the alchemical states from $\lambda=0$ to $\lambda = 1/2$ (from right to left in color sequence) showing a gap between high-energy and low-energy distributions and a bimodal distribution (pink). (C) Free energy profile along the alchemical transformation displaying a large free energy barrier.}
    \label{fig:eg5-CHEMBL1084431_CHEMBL1085666}
\end{figure}

In contrast to small R-group transformations, RBFE calculations involving a change of net charge of the ligand or a shift of a charged group from one position to another often display large uncertainties. The c-Met pair in Figure \ref{fig:cmet-CHEMBL3402745-200_CHEMBL3402743-42} where a positively charged alkylamine moiety is added to a neutral carbamate substituent, is an example of this class. While the free energy of the alchemical intermediate is high (above 40 kcal/mol), the sequence of alchemical states is relatively well connected.  The Eg5 case in Figure \ref{fig:eg5-CHEMBL1084431_CHEMBL1085666} includes a shift by three bonds of a charged ammonium group. This alchemical transformation is assigned a large uncertainty because it is characterized by both a disconnected alchemical path and a high free energy intermediate.

Overall, the 66 RBFE predictions classified as either charge-changing or charge-shifting transformations have larger uncertainties and poorer agreement with the experiments than average. This is especially so for the transformations in this class for the SHP2 set that are found to have an RMSE relative to the experiments in excess of 3 kcal/mol. Nearly all edges with large uncertainties for the c-Met and Eg5 sets (Figure \ref{fig:edges-corrplots}) correspond to charge-changing or charge-shifting transformations. Achieving better convergence for these transformations would significantly improve ATM's promising overall prediction accuracy for these sets. 

\begin{figure}
    \centering
    \includegraphics[scale = 0.25]{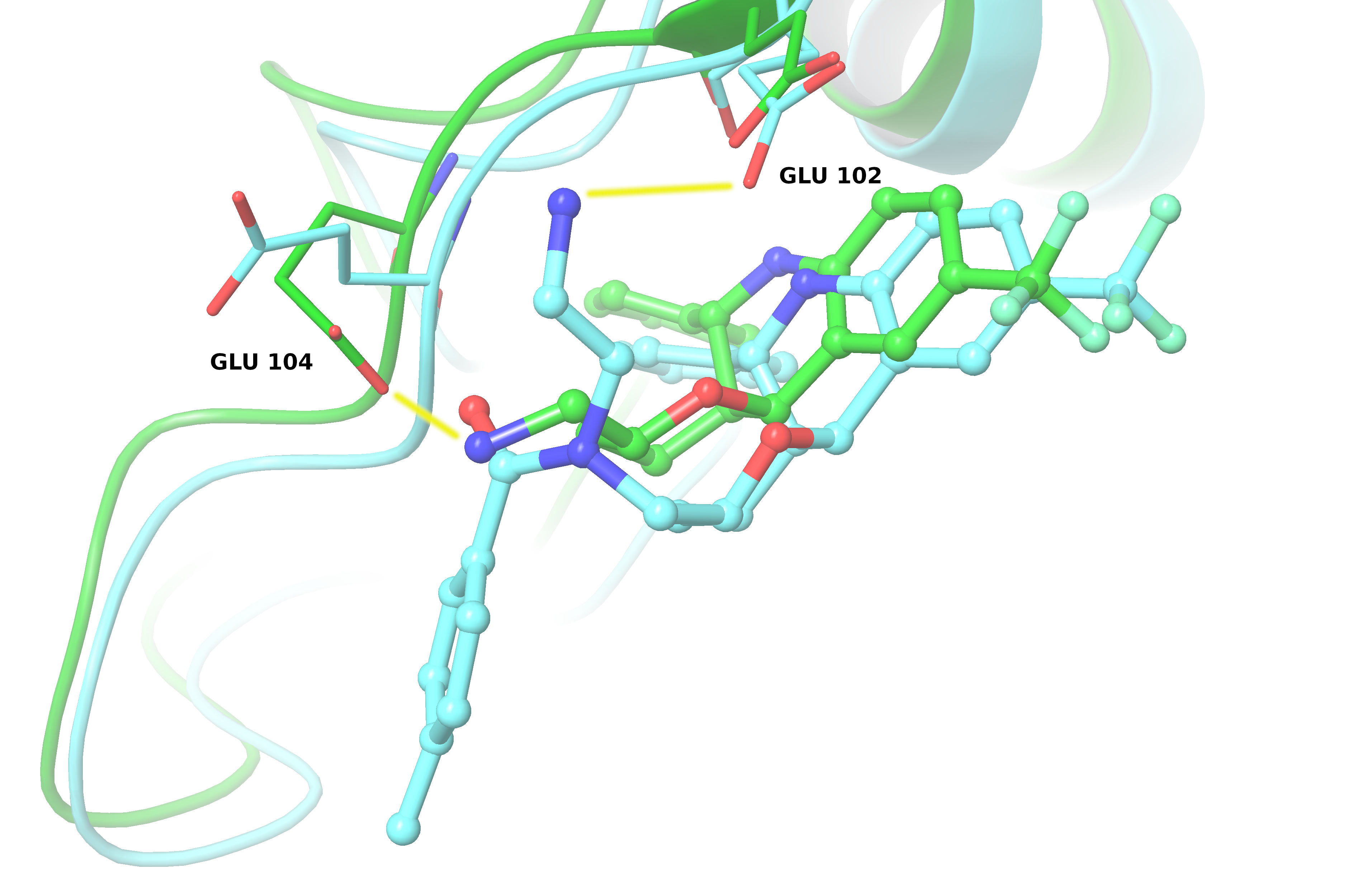}
    \caption{The structures of the complexes of CHEMBL1084431 (cyan carbon atoms, ribbon) and CHEMBL1085666 (green carbon atoms, ribbon) with the Eg5 receptor of Figure \ref{fig:eg5-CHEMBL1084431_CHEMBL1085666}. A conformational reorganization occurs whereby the salt bridge  between the ligand and the receptor (yellow bonds) switches from Glu102 to the Glu104 residue going from the first complex to the second. The structures are taken at the end of the trajectory of the alchemical end-states. } 
    \label{fig:Eg5-CHEMBL1084431_CHEMBL1085666-r3-15}
\end{figure}

Because the system remains neutral, the poorer outcomes observed here for alchemical transfer transformations involving charge variations are not obviously due to systematic boundary conditions and finite system biases present in some double-decoupling protocols.\cite{ohlknecht2020correcting,rocklin2013calculating} The convergence issues observed with charged groups are more likely related to the conformational reorganization of the complex that occurs when interactions are varied by placing or relocating a charged group of the ligand. This is illustrated by the case in Figure \ref{fig:eg5-CHEMBL1084431_CHEMBL1085666} where an amide hydrogen bonding acceptor group is replaced by a strong ammonium hydrogen bonding donor originally placed three bonds away. In this case, we observe that the two ligands shift position and form salt bridges with two different glutamate residues of the receptor (Figure \ref{fig:Eg5-CHEMBL1084431_CHEMBL1085666-r3-15}). The equilibration between these two conformational states of the complex hinders convergence because it occurs slowly relative to the timescales of the alchemical simulations. In addition, because the receptor needs to accommodate both charged groups of the ligand simultaneously, conformational frustration at the alchemical intermediate leads to high free energies and gaps and lack of state overlaps along the alchemical pathway (Figure \ref{fig:eg5-CHEMBL1084431_CHEMBL1085666}). 

This and the previous examples underscore the various ways in which differences between ligand pairs prevent successful RBFE predictions. Two ligands can have very different sizes and shapes, as in the example of Figure \ref{fig:syk-CHEMBL3265035_CHEMBL3264996}, or, as in the present example of Figure \ref{fig:eg5-CHEMBL1084431_CHEMBL1085666}, they can be structurally similar but differ radically in the way that corresponding groups interact with their environments. Cases with very dissimilar ligands such as these can probably be addressed by breaking the alchemical transformation into smaller steps by inserting suitable chemical intermediates. For example, the transformation in Figure \ref{fig:eg5-CHEMBL1084431_CHEMBL1085666} that attempts to replace one charged group with another simultaneously causing a large conformational reorganization (Figure \ref{fig:Eg5-CHEMBL1084431_CHEMBL1085666-r3-15}) could be implemented by first removing or neutralizing one ammonium group and then inserting the other in a second step. The case in Figure \ref{fig:cmet-CHEMBL3402745-200_CHEMBL3402743-42} and similar others we observed indicate that individual insertions of charged groups are more manageable than  replacing one charged group with another placed elsewhere.

\subsection{Scaffold-hopping Transformations are as Straightforward as R-group Transformations}

The Alchemical Transfer Method is based on a dual-topology representation, making it easier to set up scaffold-hopping transformations between ligands that do not share the same core topology. Indeed, the ATM setup procedure described here for scaffold-hopping transformations is the same as any other transformation. Sixty of the 548 RBFE calculations conducted in this work were classified as scaffold hopping transformations.

\begin{figure}
    \centering
    \includegraphics[scale = 0.25]{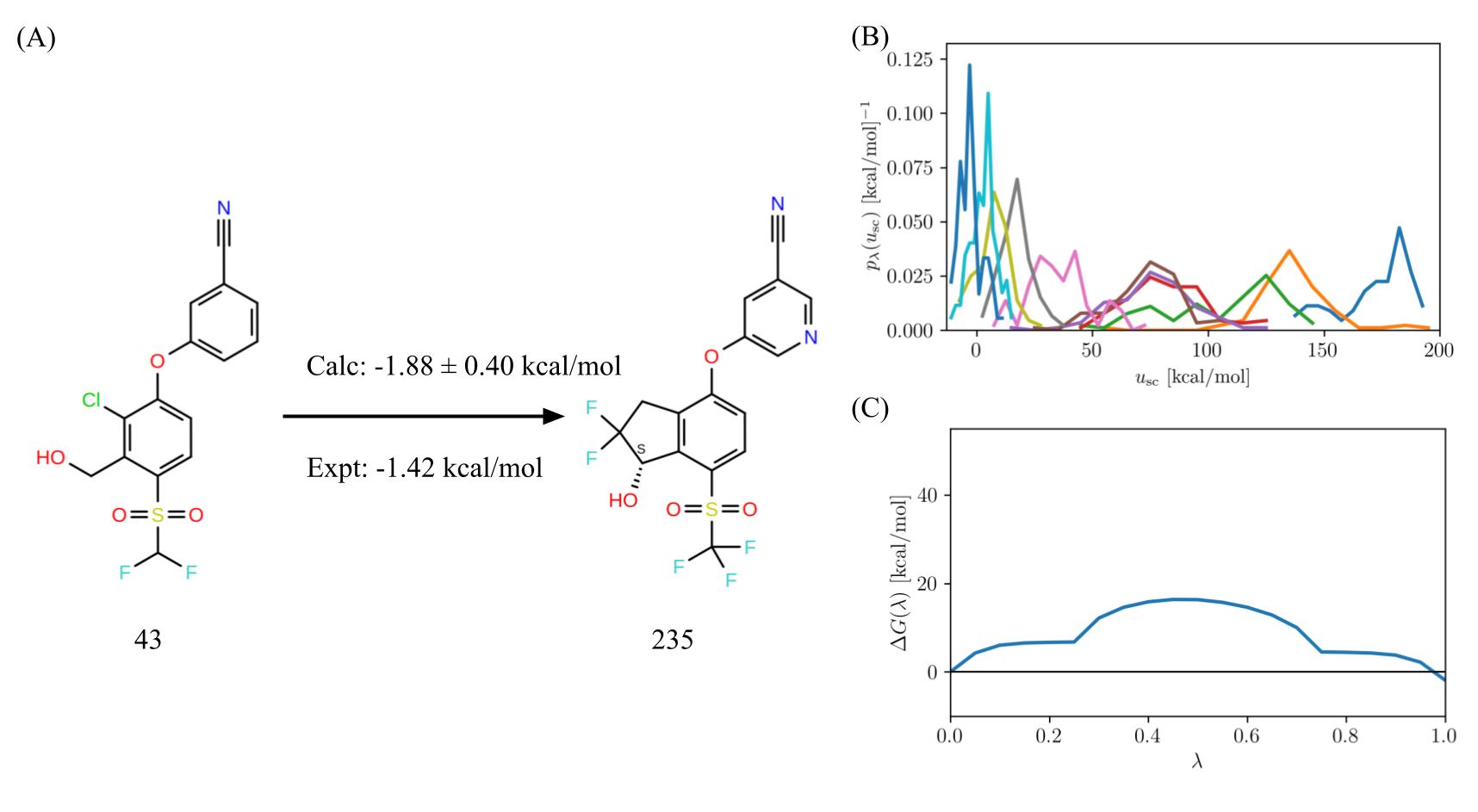}
    \caption{The results of the RBFE calculation between the ligands 43 and 235 of the HIF-2$\alpha$ receptor illustrating a reliable small scaffold-hopping cyclization transformation. (A) Chemical structures of the ligands and calculated and experimental RBFE. (B) Perturbation energy distributions for the alchemical states from $\lambda=0$ to $\lambda = 1/2$ (from right to left in color sequence) showing good overlaps between nearby distributions. (C) Free energy profile along the alchemical transformation displaying a small free energy barrier. }
    \label{fig:hif2a-43_235}
\end{figure}

\begin{figure}
    \centering
    \includegraphics[scale = 0.25]{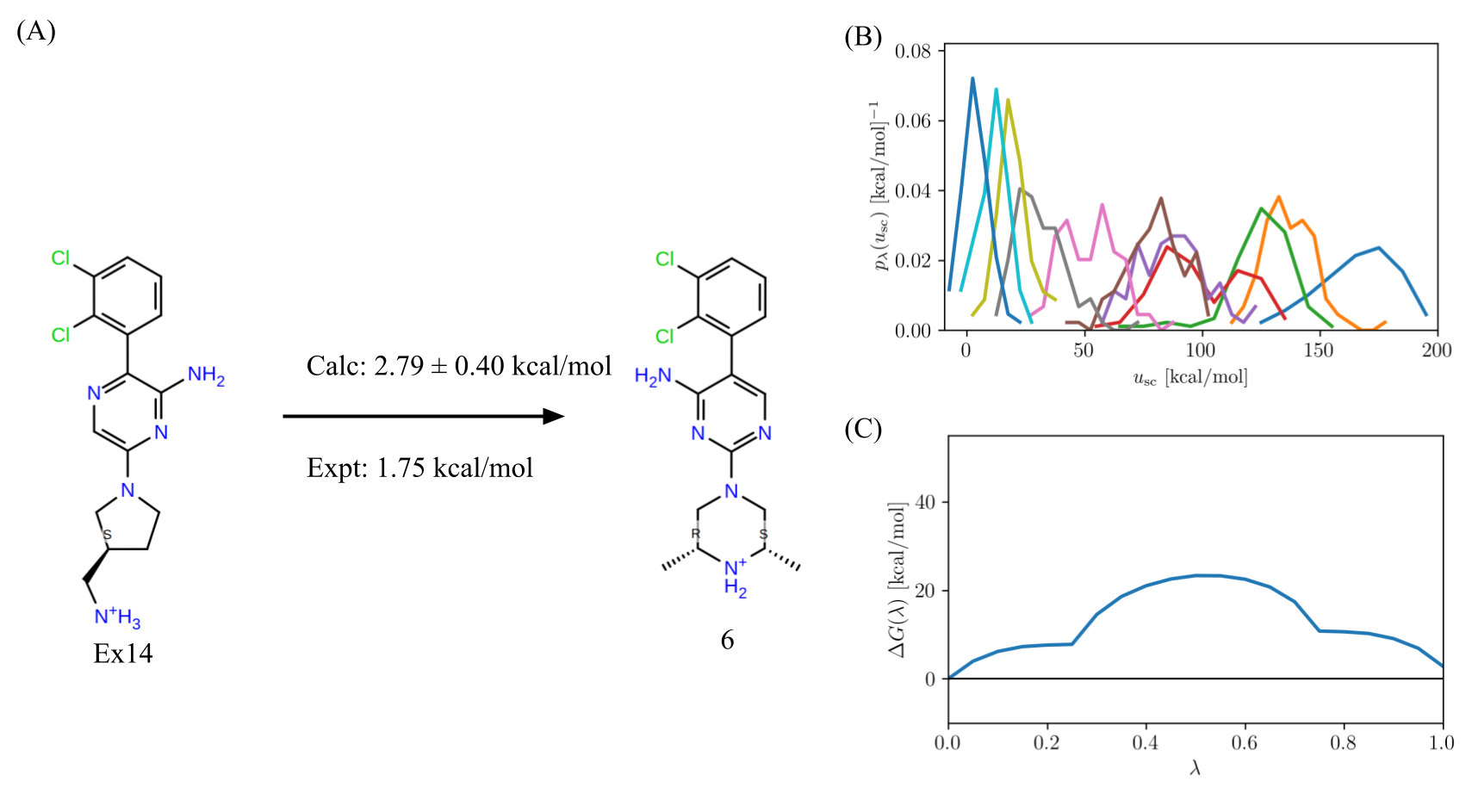}
    \caption{The results of the RBFE calculation between the ligands Example 14 and 6 of the SHP2 receptor illustrating a reliable small scaffold-hopping ring expansion transformation. (A) Chemical structures of the ligands and calculated and experimental RBFE. (B) Perturbation energy distributions for the alchemical states from $\lambda=0$ to $\lambda = 1/2$ (from right to left in color sequence) showing good overlaps between nearby distributions. (C) Free energy profile along the alchemical transformation displaying a small free energy barrier.}
    \label{fig:shp2-Ex14_6}
\end{figure}

As illustrated by the examples in Figures \ref{fig:hif2a-43_235} and \ref{fig:shp2-Ex14_6}, the results of this validation campaign have also shown that scaffold-hopping RBFE calculations converge just as efficiently as small R-group transformations of the same kind. In the HIF-2$\alpha$ case in Figure \ref{fig:hif2a-43_235}, for example, a 5-membered ring is formed by the cyclization of two substituents of the central aromatic group. In the case of Figure \ref{fig:shp2-Ex14_6}, a 5-membered ring is expanded to a 6-membered ring. In both of these cases and many similar ones we observed, there are good overlaps between perturbation energy distributions, and the free energy of the alchemical intermediate is moderate, just as in small R-group transformations (Figure \ref{fig:cmet-CHEMBL3402758-10_CHEMBL3402760-1})

\section{Conclusions}

We tested the Alchemical Transfer Method (ATM) as implemented in the open source AToM-OpenMM software package with a bespoke ligand force field on the large and challenging relative binding free energy benchmark sets developed by Schindler et al.\cite{schindler2020large} In this work, 548 RBFE calculations were set up and performed largely unsupervised using our Python workflow facilitated by the minimal ATM customization requirements. The raw RBFE estimates from ATM showed higher statistical fluctuations relative to FEP+. Nevertheless, the absolute binding free energy estimates produced by the DiffNet algorithm on this data provided an accuracy relative to the experiments comparable on average to FEP+. The results indicate that standard small R-group transformations converge rapidly, as expected. RBFE estimates for large R-group transformations and charge-changing and charge-shifting transformations are less reliable due to the slow convergence rate of conformational reorganization effects induced by the significant changes in molecular size and ligand-receptor interactions. 

Based on the insights from this work, we plan to break up difficult RBFE transformations into multiple, more manageable steps. For example, large R-group transformations will be implemented as a series of smaller transformations involving suitable chemical intermediates. Similarly, to minimize the impact of conformational reorganization effects, we will consider breaking up transformations involving the replacement of an exclusive hydrogen bond donor with an exclusive hydrogen bond acceptor using chemical intermediates containing groups, such as hydroxyls, with promiscuous hydrogen bond behavior. This study further confirms that scaffold-hopping transformations involving the formation or breaking of chemical bonds do not provide additional challenges with this method. In conclusion, this study confirms that ATM is a promising production tool for lead optimization in structure-based drug discovery. 

\section{Acknowledgments}
We acknowledge support from the National Science Foundation (NSF CAREER 1750511 to E.G.). We are grateful for the computer time provided on Roivant's Neo computational cluster.  



\providecommand{\latin}[1]{#1}
\makeatletter
\providecommand{\doi}
  {\begingroup\let\do\@makeother\dospecials
  \catcode`\{=1 \catcode`\}=2 \doi@aux}
\providecommand{\doi@aux}[1]{\endgroup\texttt{#1}}
\makeatother
\providecommand*\mcitethebibliography{\thebibliography}
\csname @ifundefined\endcsname{endmcitethebibliography}
  {\let\endmcitethebibliography\endthebibliography}{}

\end{document}